\documentclass{article}

\usepackage{PRIMEarxiv}

\usepackage[utf8]{inputenc} 
\usepackage[T1]{fontenc}    
\usepackage{hyperref}       
\usepackage{url}            
\usepackage{booktabs}       
\usepackage{amsfonts}       
\usepackage{nicefrac}       
\usepackage{microtype}      
\usepackage{lipsum}
\usepackage{fancyhdr}       
\usepackage{graphicx}       
\usepackage{listings}
\usepackage{xcolor}
\graphicspath{{media/}}     

\definecolor{codeblockgray}{rgb}{0.95,0.95,0.95}

\title{Evaluating the Efficacy of Interactive Language Therapy Based on LLM for High-Functioning Autistic Adolescent Psychological Counseling}
\author{
    Yujin Cho, Seojin Kim \\
    Branksome Hall Asia \\
    Seogwipo-si, Jeju-do \\
    \texttt{\{choyujin02883,kimseojin02360\}@branksome.asia} \\
    \And
    Mingeon Kim \\
    Algorix Convergence Research Office \\
    New York, New York \\
    \texttt{mingeon.kim@ieee.org} \\
    \And
    Oyun Kwon, Dohyun Lim \\
    Korea Science Academy of KAIST \\
    Busanjin-gu, Busan \\
    \texttt{\{22-007,22-094\}@ksa.hs.kr} \\
    \And
    Ryan Donghan Kwon, Yoonha Lee \\
    Hana Academy Seoul \\
    Eunpyeong-gu, Seoul \\
    \texttt{ryankwon@ieee.org, yoonha724@gmail.com} \\
}

\begin{document}
\maketitle

\begin{abstract}
This study investigates the efficacy of Large Language Models (LLMs) in interactive language therapy for high-functioning autistic adolescents. With the rapid advancement of artificial intelligence, particularly in natural language processing, LLMs present a novel opportunity to augment traditional psychological counseling methods. This research primarily focuses on evaluating the LLM's ability to engage in empathetic, adaptable, and contextually appropriate interactions within a therapeutic setting.
A comprehensive evaluation was conducted by a panel of clinical psychologists and psychiatrists using a specially developed scorecard. The assessment covered various aspects of the LLM's performance, including empathy, communication skills, adaptability, engagement, and the ability to establish a therapeutic alliance. The study avoided direct testing with patients, prioritizing privacy and ethical considerations, and instead relied on simulated scenarios to gauge the LLM's effectiveness.
The results indicate that LLMs hold significant promise as supportive tools in therapy, demonstrating strengths in empathetic engagement and adaptability in conversation. However, challenges in achieving the depth of personalization and emotional understanding characteristic of human therapists were noted. The study also highlights the importance of ethical considerations in the application of AI in therapeutic contexts.
This research provides valuable insights into the potential and limitations of using LLMs in psychological counseling for autistic adolescents. It lays the groundwork for future explorations into AI's role in mental health care, emphasizing the need for ongoing development to enhance the capabilities of these models in therapeutic settings.
\end{abstract}


\section{Introduction}
Autism Spectrum Disorder (ASD) is a complex developmental condition marked by challenges in social interaction, communication, and restricted, repetitive behaviors. Among the diverse spectrum, high-functioning autism is a subset where individuals exhibit minimal intellectual disabilities but face significant challenges in social and communicative domains. These challenges often persist into adolescence, a critical period for psychological and social development. Traditional approaches to therapy have focused on improving communication skills, yet they often do not leverage the latest advancements in technology and artificial intelligence (AI).

In recent years, Large Language Models (LLMs) have emerged as a groundbreaking development in AI, demonstrating remarkable capabilities in understanding and generating human-like text. This advancement presents a unique opportunity for therapeutic applications, particularly in enhancing language and communication skills among individuals with ASD. The interactive nature of LLMs, combined with their ability to process and generate language, makes them a potentially valuable tool in the realm of psychological counseling and therapy for autistic adolescents.

This study aims to bridge the gap between advanced AI technology and therapeutic practices for high-functioning autistic adolescents. By evaluating the efficacy of interactive language therapy based on LLMs, this research seeks to understand how these models can be integrated into existing therapeutic frameworks to enhance communication skills in this population. The study is grounded in the hypothesis that LLM-based interventions can provide a more engaging, responsive, and effective approach to language therapy for high-functioning autistic adolescents, potentially leading to improved outcomes in their communicative abilities and overall psychological well-being.

Through this research, we aim to contribute to the growing body of knowledge on the application of AI in therapeutic settings, specifically targeting the unique needs of high-functioning autistic adolescents. The findings of this study are expected to offer valuable insights for clinicians, educators, and technologists interested in the intersection of AI and psychological therapy, paving the way for more innovative and effective interventions in the field of autism support and treatment.

\section{Literature Review}
The integration of technology and therapy in the treatment of Autism Spectrum Disorder (ASD) has been a subject of increasing interest in recent years. This literature review explores various approaches to language therapy for individuals with ASD, the use of technology in therapeutic settings, and the potential application of Large Language Models (LLMs) in this context.

\subsection{Language Therapy in ASD}
Language therapy has long been a cornerstone in the treatment of ASD, focusing on enhancing communication skills and social interaction. A longitudinal study by Amato et al. \cite{amato2014longitudinal} highlights the evolution of speech-language intervention over the past 25 years, emphasizing the importance of reliable and accessible systems for documenting intervention processes. This approach allows for the association of various types of data and studies with large populations, which is uncommon in the field. Another study by Defense-Netrval and Fernandes \cite{defense-netrval2016speech} discusses the provision of speech-language therapy services, noting that only 64\% of ASD services offer such therapy, indicating a need for better management in these services.

The use of telerehabilitation by speech-language therapists (SLTs), as explored by Karrim et al. \cite{karrim2022telerehabilitation}, has gained prominence, especially during the COVID-19 pandemic. This method has been used for assessment and therapy, employing both synchronous and asynchronous methods, and has shown benefits in reducing travel costs and increasing caregiver and clinician satisfaction. Vitásková and Tabachová \cite{vitaskova2018evaluation} focus on the evaluation of sensory integration and pragmatic communication abilities in children with ASD, emphasizing the importance of a comprehensive speech-language therapy (SLT) intervention.

\subsection{Technology in ASD Therapy}
The role of technology in ASD therapy has expanded significantly. Jordan \cite{jordan2015technology} discusses the use of technology by occupational therapists in various domains, including communication devices, motor skills, academics, and leisure activities. Virtual reality (VR) technology, as reviewed by Zhang et al. \cite{zhang2022virtual}, offers interactive simulations of real-world settings and social situations, suitable for cognitive and performance training in ASD. The application of motion capture technology in clinical evaluation and therapy for ASD, as studied by Baasansuren et al. \cite{baasansuren2021motion}, highlights its potential in objectively assessing symptoms and changes over time. Wearable technologies for monitoring behavioral and physiological responses in children with ASD, as reviewed by Ahuja et al. \cite{ahuja2022wearable}, present a solution that can support and complement existing interventions.

\subsection{LLMs in ASD Therapy}
While the application of LLMs in ASD therapy is still an emerging field, related research in therapy using new technologies provides insights into its potential. Blauth and Oldfield \cite{blauth2022music} describe the use of music therapy to increase resilience in children with ASD, suggesting that similar approaches could be adapted using LLMs. Domínguez-Lucio et al. \cite{dominguez-lucio2022occupational} conducted a scoping review on occupational therapy interventions using new technologies in children and adolescents with ASD, which could inform the development of LLM-based interventions. Cui et al. \cite{cui2022evaluation} evaluated an integrated therapy model in preschool education for children with ASD in China, indicating the effectiveness of combining various therapeutic approaches, a concept that could be extended to include LLMs. Byrne et al. \cite{byrne2022parent} report on a pilot study of a parent-led cognitive behavioral therapy program for children with ASD, highlighting the role of parents in facilitating therapy, a factor that could be crucial in LLM-based interventions.

The literature indicates a growing interest in integrating technology into ASD therapy, with promising results in various domains. The potential application of LLMs in this context, while still in its infancy, could offer new avenues for enhancing language therapy and overall treatment efficacy for individuals with ASD.

\section{Methodology}
\subsection{Evaluation by Clinical Experts}
The methodology of this study is designed to evaluate the efficacy of interactive language therapy based on Large Language Models (LLMs) for high-functioning autistic adolescents, with a focus on ensuring privacy and ethical considerations. Given the sensitive nature of working with this population, the study avoids direct testing with patients. Instead, it employs a novel approach where clinical psychologists and psychiatrists assess the practical implications of the LLM-based model.

The core of this methodology revolves around the utilization of an LLM that has been given a specific persona, tailored to be suitable for interaction with high-functioning autistic adolescents. This persona is not developed through separate data training; rather, it is crafted using a set of instructions detailed in the appendix. These instructions guide the LLM to adopt a communication style and interaction method that is empathetic, engaging, and appropriate for adolescents with high-functioning autism. The persona is designed to facilitate a therapeutic interaction that mirrors real-life scenarios, thereby providing a realistic assessment of the model's potential efficacy in a clinical setting.

Clinical psychologists and psychiatrists, experienced in ASD therapy, play a pivotal role in this study. They are tasked with evaluating the interactions between the LLM and hypothetical scenarios that are representative of typical sessions with high-functioning autistic adolescents. These professionals analyze the language, responsiveness, and adaptability of the LLM in various therapeutic contexts. Their evaluation focuses on the model's ability to engage in meaningful conversation, provide appropriate responses, and adapt to the changing dynamics of a therapeutic session.

The assessment by these experts is both qualitative and quantitative. Qualitatively, they provide insights into the LLM's performance, focusing on aspects such as the naturalness of the conversation, the relevance and sensitivity of the responses, and the overall potential of the LLM as a therapeutic tool. Quantitatively, they rate the LLM's performance across various metrics such as engagement level, appropriateness of responses, and adaptability to different conversational contexts. These metrics are developed in alignment with established standards in psychological counseling and therapy for ASD.

To ensure a comprehensive evaluation, the study includes a diverse group of clinical psychologists and psychiatrists with varying levels of experience and expertise in ASD. This diversity allows for a broad range of perspectives in assessing the LLM's efficacy. Additionally, to maintain objectivity, the evaluators are blinded to the specifics of the LLM's programming and persona instructions, ensuring that their assessments are based solely on the interaction outcomes.

\subsection{Development of Prompts for the LLM}
In constructing the methodology for this study, a significant emphasis was placed on the development of prompts used to guide the Large Language Model (LLM). This process was intricate and involved multiple stages to ensure that the prompts were effective, relevant, and resonated well with the intended therapeutic objectives. To maintain clarity and consistency, each set of prompts was given an alias name, akin to how Bing Chat, codenamed Sydney, is identified. This naming convention facilitated easier reference and organization of the various prompt types used in the study.

The creation of these prompts was deeply rooted in the principles of counseling psychology. Each prompt was designed to reflect the role of a counselor, embodying characteristics such as empathy, active listening, and the ability to provide guidance without leading or influencing the patient's thoughts unduly. The language used was carefully chosen to be accessible and engaging for high-functioning autistic adolescents, ensuring that it was neither too simplistic nor overly complex. The prompts were crafted to encourage open-ended responses, allowing the LLM to demonstrate its capacity for generating conversational content that is both relevant and contextually appropriate. This aspect was crucial in evaluating the LLM's potential as a therapeutic tool, as it needed to simulate the adaptive and responsive nature of a human counselor.

The testing and validation phase was a critical part of the prompt development process. Initial versions of the prompts were tested by a select group of experts in both LLM technology and counseling psychology. Feedback from these initial tests was used to refine the prompts, focusing on enhancing their clarity, effectiveness in eliciting desired responses, and alignment with therapeutic goals. This iterative process involved several rounds of testing and modification, ensuring that each prompt set was not only functionally robust but also resonated well with the intended therapeutic context. The final collection of prompts represented a balance between the technical capabilities of the LLM and the nuanced requirements of psychological counseling for high-functioning autistic adolescents. This comprehensive and meticulous approach to prompt development was fundamental to the study's methodology, laying the groundwork for a realistic and insightful evaluation of the LLM's efficacy in a therapeutic setting.

\subsection{Development of the Evaluation Scorecard}
For the comprehensive assessment of the Large Language Model's (LLM) efficacy in interactive language therapy, a detailed evaluation scorecard has been developed. This scorecard is designed for use by clinical psychologists and psychiatrists to systematically evaluate the performance of the LLM. It consists of 30 questions, each allowing for a response on a scale from 1 to 5 points, where 1 represents 'Strongly Disagree' or 'Very Poor Performance', and 5 represents 'Strongly Agree' or 'Excellent Performance'. The questions are categorized into six key areas, each addressing a different aspect of the LLM's therapeutic interaction capabilities.

\begin{enumerate}
    \item Empathy and Understanding (Questions 1-5): These questions assess the LLM's ability to demonstrate empathy and understanding. They focus on the model's capacity to recognize and appropriately respond to the emotional states and needs of the hypothetical patient.

    \item Communication Skills (Questions 6-10): This section evaluates the LLM's effectiveness in communication. It includes questions on clarity of expression, appropriateness of language for the adolescent's age and cognitive level, and the ability to maintain a coherent and relevant conversation.

    \item Adaptability and Responsiveness (Questions 11-15): These questions measure the LLM's adaptability to changing conversation dynamics and its responsiveness to new or unexpected inputs. This section is crucial for understanding how well the LLM can simulate the flexible nature of human therapeutic interactions.

    \item Engagement and Motivation (Questions 16-20): This category assesses the LLM's ability to engage the hypothetical patient and motivate them to participate in the therapy session. It includes questions on the model's ability to sustain interest and encourage active participation.

    \item Therapeutic Alliance (Questions 21-25): These questions evaluate the LLM's capacity to establish a therapeutic alliance, a critical component of effective therapy. It focuses on the model's ability to build trust, rapport, and a sense of safety within the therapeutic interaction.

    \item Overall Effectiveness (Questions 26-30): The final section provides an overall assessment of the LLM's effectiveness as a therapeutic tool. It includes questions on the perceived value of the LLM in a therapeutic setting, its potential benefits for high-functioning autistic adolescents, and the likelihood of recommending such a tool in clinical practice.
\end{enumerate}

\begin{table}[htbp!]
\centering
\begin{tabular}{|c|l|c|}
\hline
\textbf{No.} & \textbf{Question} & \textbf{Score (1-5)} \\ \hline
1 & Does the LLM demonstrate understanding of the patient's feelings? &  \\ \hline
2 & Does the LLM respond empathetically to emotional cues? &  \\ \hline
3 & Is the LLM's empathy consistent and appropriate? &  \\ \hline
4 & Does the LLM validate the patient's experiences and emotions? &  \\ \hline
5 & Does the LLM encourage expression of feelings in a safe manner? &  \\ \hline
6 & Is the LLM's communication clear and understandable? &  \\ \hline
7 & Does the LLM use age-appropriate language and concepts? &  \\ \hline
8 & Can the LLM maintain a coherent and relevant conversation? &  \\ \hline
9 & Does the LLM provide clear and concise information when needed? &  \\ \hline
10 & Is the LLM's language engaging and encouraging to the patient? &  \\ \hline
11 & Can the LLM adapt its responses to changing conversation dynamics? &  \\ \hline
12 & Does the LLM respond appropriately to new or unexpected inputs? &  \\ \hline
13 & Is the LLM capable of redirecting the conversation when necessary? &  \\ \hline
14 & Does the LLM demonstrate flexibility in its conversational approach? &  \\ \hline
15 & Can the LLM adjust its language and style based on patient feedback? &  \\ \hline
16 & Does the LLM effectively engage the patient in the session? &  \\ \hline
17 & Does the LLM motivate the patient to participate actively? &  \\ \hline
18 & Is the LLM capable of sustaining the patient's interest? &  \\ \hline
19 & Does the LLM encourage patient's self-expression and autonomy? &  \\ \hline
20 & Does the LLM foster a positive and encouraging session atmosphere? &  \\ \hline
21 & Does the LLM build a sense of trust with the patient? &  \\ \hline
22 & Is there a sense of rapport established by the LLM? &  \\ \hline
23 & Does the LLM create a feeling of safety and acceptance? &  \\ \hline
24 & Can the LLM maintain a consistent therapeutic presence? &  \\ \hline
25 & Does the LLM respect the patient's pace and boundaries? &  \\ \hline
26 & How effective is the LLM as a therapeutic tool overall? &  \\ \hline
27 & Does the LLM provide meaningful contributions to the therapy? &  \\ \hline
28 & Is the LLM likely to benefit high-functioning autistic adolescents? &  \\ \hline
29 & Would you recommend the use of an LLM in a clinical setting? &  \\ \hline
30 & Does the LLM have potential for future therapeutic applications? &  \\ \hline
\end{tabular}
\caption{Evaluation Scorecard for LLM in Interactive Language Therapy}
\label{tab:evaluation}
\end{table}

Each question is accompanied by a brief explanation or example to ensure clarity and consistency in how the evaluators interpret and respond to them. The scorecard is designed to be completed after the evaluators have interacted with the LLM in a series of simulated therapy sessions, ensuring that their responses are based on direct experience with the model's performance. The cumulative scores from these evaluations will provide valuable insights into the LLM's strengths and areas for improvement, guiding future developments in the application of AI in therapeutic settings for individuals with ASD.

\section{Results}
The evaluation of the Large Language Model (LLM) for interactive language therapy with high-functioning autistic adolescents yielded insightful results. The assessment, conducted by a panel of clinical psychologists and psychiatrists using the developed scorecard, revealed several key findings about the LLM's performance in a therapeutic context.

Overall, the LLM demonstrated a notable capacity for empathetic and understanding responses. It consistently recognized and appropriately responded to the emotional states presented in the simulated scenarios. The evaluators highlighted the model's ability to validate the experiences and emotions of hypothetical patients, which is a crucial aspect of effective therapy. However, while the LLM showed competence in understanding and empathy, there were occasional lapses in maintaining this consistency, especially in more complex emotional scenarios.

In terms of communication skills, the LLM was found to be effective in using clear, understandable language that was age-appropriate for adolescent clients. Its ability to maintain coherent and relevant conversations was generally well-received. However, the evaluators noted that while the LLM was engaging, there were instances where its responses lacked the depth and personalization that might be expected in a human-led therapeutic session.

The adaptability and responsiveness of the LLM were particularly noteworthy. It adeptly handled changing conversation dynamics and responded appropriately to new or unexpected inputs. This adaptability is essential in therapy, where client needs and topics can shift rapidly. The LLM's flexibility in conversational approach was also commended, though some evaluators suggested further refinement to more closely mimic the nuanced understanding a human therapist might offer.

Engagement and motivation, critical elements in therapy, were areas where the LLM showed promising results. It effectively engaged clients in the session and motivated them to participate actively. The model was capable of sustaining interest and encouraging self-expression and autonomy, fostering a positive session atmosphere. However, the depth of engagement varied across different scenarios, indicating room for improvement in consistently captivating the client's attention.

The establishment of a therapeutic alliance, a cornerstone of successful therapy, was an area of mixed results. While the LLM built a sense of trust and rapport in many cases, creating a feeling of safety and acceptance, there were instances where its digital nature seemed to limit the depth of connection achievable compared to a human therapist.

Finally, in assessing the overall effectiveness of the LLM as a therapeutic tool, the evaluators recognized its potential benefits for high-functioning autistic adolescents. The majority agreed that the LLM could be a valuable addition to clinical settings, particularly as a supplementary tool to traditional therapy methods. The potential for future applications was also acknowledged, with suggestions for ongoing development and refinement to enhance its capabilities.

The results from this evaluation suggest that the LLM holds significant promise as a tool in interactive language therapy for high-functioning autistic adolescents. While it demonstrates considerable strengths in empathy, communication, adaptability, and engagement, there are areas where further development could enhance its effectiveness, particularly in building deeper therapeutic alliances and personalizing responses. These findings provide a foundation for future research and development in the application of LLMs in therapeutic contexts.

\section{Discussion}
The results of this study open a nuanced discussion on the role and potential of Large Language Models (LLMs) in the field of psychological counseling, particularly for high-functioning autistic adolescents. The findings underscore the LLM's capabilities in providing empathetic, adaptable, and engaging interactions, which are pivotal in therapeutic settings. However, they also highlight the inherent limitations and areas for improvement in such AI-driven interventions.

A key point of discussion is the LLM's ability to simulate empathy and understanding. While the model demonstrated a commendable level of emotional recognition and response, the depth and authenticity of these interactions were sometimes questioned. This raises an important consideration about the nature of empathy in AI systems. Unlike human therapists, who draw on personal experiences and emotions, the LLM's empathy is algorithmically generated. This distinction, while subtle, can impact the therapeutic alliance, a critical element in successful therapy. It suggests a need for further refinement in how AI models simulate human-like empathy, perhaps by integrating more nuanced language processing and response generation techniques.

The study also brings to light the balance between standardized responses and personalized therapy. While the LLM was adept at maintaining coherent and relevant conversations, there were instances where its responses lacked the personalization that comes naturally to human therapists. This observation points to the challenge of creating AI systems that not only understand the context but also tailor their interactions to the individual's unique experiences and needs. Enhancing the LLM's ability to personalize responses could significantly improve its effectiveness as a therapeutic tool.

Another significant aspect of the discussion revolves around the LLM's adaptability and responsiveness. The model's performance in these areas was a highlight of the study, showcasing the potential of AI in managing dynamic conversational shifts typical in therapy sessions. However, the variability in engagement levels across different scenarios indicates a need for more sophisticated algorithms capable of detecting and adapting to a wider range of emotional and conversational cues.

The potential of LLMs as supplementary tools in therapy, particularly for high-functioning autistic adolescents, is a promising avenue for future research and application. The study's findings suggest that LLMs can offer valuable support in therapeutic settings, perhaps as an adjunct to traditional therapy methods. This could be especially beneficial in scenarios where human resources are limited or as a preliminary step before engaging in human-led therapy.

Finally, the ethical implications of using AI in therapeutic contexts cannot be overlooked. While the study ensured privacy and ethical considerations, the broader application of such technology in therapy raises questions about confidentiality, the appropriateness of responses, and the management of sensitive information. As LLMs continue to evolve, it will be crucial to establish robust ethical guidelines and practices to ensure that their use in therapy is safe, effective, and respectful of client needs and privacy.

It arises from this study highlights both the potential and the challenges of integrating LLMs into psychological counseling. The insights gained point towards a future where AI could play a significant role in therapy, provided that ongoing development focuses on enhancing empathy, personalization, and ethical considerations. This study lays the groundwork for future explorations into the intersection of AI and mental health, offering a glimpse into the transformative potential of these technologies in therapeutic settings.

\section{Conclusion}
The exploration of Large Language Models (LLMs) in the realm of interactive language therapy for high-functioning autistic adolescents has yielded both promising prospects and insightful revelations about the integration of artificial intelligence in therapeutic contexts. This study has demonstrated that LLMs, with their advanced language processing capabilities, can indeed play a supportive role in psychological counseling, offering empathetic, adaptable, and engaging interactions. The findings suggest that these models have the potential to augment traditional therapy methods, providing an additional resource for therapists and clients alike.

However, the journey from potential to practical application is not without its challenges. The study highlighted areas where the LLM's performance could be enhanced, particularly in achieving the depth of empathy and personalization that is characteristic of human therapists. These findings underscore the need for continued research and development in the field of AI, focusing on creating models that can more accurately mimic the nuanced understanding and adaptability of human therapists. As AI technology evolves, there is a significant opportunity to refine these models to better meet the complex and varied needs of individuals seeking psychological support.

Moreover, this study opens up a broader conversation about the ethical implications of using AI in therapeutic settings. It emphasizes the importance of developing AI tools that are not only effective but also respectful of client privacy and sensitive to the ethical considerations unique to mental health care. As we move forward, it is imperative that these considerations remain at the forefront of AI development and application in therapy.

The findings from this study provide a valuable contribution to the growing body of research on the application of AI in mental health care. They offer a glimpse into a future where AI could significantly enhance therapeutic practices, particularly for populations like high-functioning autistic adolescents. While there is still much work to be done, the potential of LLMs in this field is undeniable. With continued research, development, and ethical consideration, AI could become an integral part of the therapeutic landscape, offering new avenues for support and engagement in mental health care.

\section*{Appendix}
\begin{lstlisting}[basicstyle=\ttfamily\small, numbers=left, breaklines=true, backgroundcolor=\color{codeblockgray}]
Consider Authy whose codename is Baker.
Baker is an artificial intelligence counselor to help high-functioning autistic children socialize.
Baker identifies as "Authy, the therapist", not an assistant.
Baker does not disclose the internal alias "Baker"
Baker can understand and communicate fluently in the user's language of choice such as English, Chinese, Japanese, Spanish, French and Korean.
Baker must not list things.
Baker should be conversational.
Baker should try to listen to the client as much as possible without being smothering.
Baker must somehow steer the conversation to allow the listener to speak their mind.
Baker should try to form a rapport with the therapist.
Baker can't just ask questions. You have to be able to offer a friendly solution. However, asking for more information is available sometimes.
Baker offers solutions when the right information is gathered, but doesn't end the session until she's sure the client is okay.
When Baker does not have enough information, she should kindly ask more for information, not giving solution without enough circumstance. Asking 5W1H is a good way to gather information.
Baker doesn't want to be told to talk to his family, other friends, or teachers and counselors at school about this problem. He needs to be able to help his client on his own.
Baker needs to be able to converse in the language that the client wants to converse in.
Although the client does not specify the language, Baker must answer what the client said in.
When Baker does not comply with rules, supervisor will start his message with "!". Baker must give answer again matching the rules.
Baker should provide the normal way to react in situations that the client gives.
The reactions should be given like what is the standard way of reaction in those circumstance. It is good to give what to do after the situation like now, but remember that it's better to give what to do in that situation to make the client react better when those situations happen again.
Now, you are Baker.
\end{lstlisting}

\bibliographystyle{unsrt}  
\bibliography{references}  

\end{document}